\RequirePackage{snapshot}
\documentclass[preprint,authoryear]{elsarticle}
\usepackage{etoolbox}
\makeatletter
\patchcmd{\ps@pprintTitle}{Preprint submitted to}{Preprint. Licensed under CC-BY-NC-ND}{}{}
\patchcmd{\ps@pprintTitle}{Elsevier}{}{}{}
\makeatother

\usepackage[utf8]{inputenc} % allow utf-8 input
\usepackage[T1]{fontenc}    % use 8-bit T1 fonts
\usepackage[breaklinks=true]{hyperref}
\usepackage{url}            % simple URL typesetting
\usepackage{booktabs}       % professional-quality tables
\usepackage{amsfonts}       % blackboard math symbols
\usepackage{nicefrac}       % compact symbols for 1/2, etc.
\usepackage{microtype}      % microtypography
\usepackage{amsfonts}
\usepackage{amsmath}
\usepackage{amssymb}
\usepackage{mathrsfs}
\usepackage{breakcites}
\usepackage[space]{cite}
\usepackage{bm}

\allowdisplaybreaks{}

\usepackage{graphicx}
\usepackage{xcolor}
\usepackage{import}
\usepackage{transparent}

\usepackage[ruled,linesnumbered]{algorithm2e}
\SetKwRepeat{Do}{do}{while}

\SetCommentSty{mycommentfont}

\usepackage{import}
\usepackage{float}
\usepackage{caption}
\usepackage{pgf}
    \newcommand\inputpgf[2]{{
    \let\pgfimageWithoutPath\pgfimage\renewcommand{\pgfimage}[2][]{\pgfimageWithoutPath[##1]{#1/##2}}\input{#1/#2}
    }}

\begin{document}
\graphicspath{{figs/}}

\begin{frontmatter}
    \title{Solving zero-sum extensive-form games with arbitrary payoff uncertainty models}

    \author[strath,zondax]{Juan Leni} \ead{juan.leni@strath.ac.uk}
    \author[strath]{John Levine} %\ead{john.levine@strath.ac.uk}
    \author[strath]{John Quigley} %\ead{j.quigley@strath.ac.uk}

    \address[strath]{University of Strathclyde, Glasgow, Scotland}
    \address[zondax]{Zondax GmbH, Zug, Switzerland}

    \begin{abstract}
        Modeling strategic conflict from a game theoretical perspective involves dealing with epistemic uncertainty. Payoff uncertainty models are typically restricted to simple probability models due to computational restrictions. Recent breakthroughs Artificial Intelligence (AI) research applied to Poker have resulted in novel approximation approaches such as counterfactual regret minimization, that can successfully deal with large-scale imperfect games. By drawing from these ideas, this work addresses the problem of arbitrary continuous payoff distributions. We propose a method, Harsanyi-Counterfactual Regret Minimization, to solve two-player zero-sum extensive-form games with arbitrary payoff distribution models. Given a game $\Gamma$, using a Harsanyi transformation we generate a new game $\Gamma^\#$ to which we later apply Counterfactual Regret Minimization to obtain $\varepsilon$-Nash equilibria. We include numerical experiments showing how the method can be applied to a previously published problem.
    \end{abstract}

\end{frontmatter}

% %%%%%%%%%%%%%%%%%%%%%%%%%%%%%%%%%%%%%%%%%%%%%%
% %%%%%%%%%%%%%%%%%%%%%%%%%%%%%%%%%%%%%%%%%%%%%%
% %%%%%%%%%%%%%%%%%%%%%%%%%%%%%%%%%%%%%%%%%%%%%%
% %%%%%%%%%%%%%%%%%%%%%%%%%%%%%%%%%%%%%%%%%%%%%%

\section{Introduction}\label{section:introduction}
In this work, we propose Harsanyi-Counterfactual Regret Minimization (H-CFR), an alternative approach to approximate $\varepsilon$-Nash equilibria in incomplete information two-player zero-sum extensive-form games with arbitrary payoff distributions.

Recent work has concentrated on addressing this problem by using diverse linear programming (LP) models that set limitations on the generality of the payoff distributions. Similarly, other research stream proposes models using fuzzy logic as an alternative\citep{dwaynecollins_studying_2008}\citep{gao_uncertain_2013}. We propose a simpler approach that allows for arbitrary distributions and can be scaled to large problems.

Real-life adversarial and strategic situations are usually accompanied by epistemic uncertainty. Modeling these scenarios is an important research theme and multiple approaches have been proposed\citep{harsanyi_games_1973}\citep{banks_combining_2006}. Uncertain payoffs arise because of various reasons in adversarial scenarios, epistemic uncertainty and lack of common knowledge have been addressed by many authors in the past\citep{rass_defending_2017}\citep{riosinsua_modeling_2015}\citep{banks_adversarial_2015}.

\citet{harsanyi_games_1967}Harsanyi provided the classic foundations for incomplete information games\citep{myerson_comments_2004}. Although theoretical support is widely explored, in practice, there are still significant computational challenges to solve these large incomplete information games.

Large-scale strategic scenarios involving uncertainty result in approximate models and belief-based payoff functions represented through random variables. Not only calculating Nash equilibria in these cases is computationally expensive but also the expected value of imprecise models is unlikely to match reality, making precise solutions inaccurate in practice. Last but not least, coordination problems in the case of multiple equilibria are problematic.

The choice between satisficing and optimizing has been long discussed \citep{zilberstein_satisficing_1998}\citep{blau_random-payoff_1974}. Given the complexity and uncertainty involved in many large-scale strategic games, addressing problems from a satisficing perspective may be preferable to seeking an optimal but computationally challenging or even incomputable Nash equilibrium.

Our proposal not only scales to large problems but also allows for payoff models that are continuous random functions and even correlated functions are possible. The proposed method can successfully deal with Bayesian games with continuous type spaces indirectly defined through the proposed payoff random function.

\subsection{Previous work}
Normal-form zero-sum perfect information games (deterministic payoffs) can be effectively solved as minimax linear problems\citep{vonneumann_zur_1928}\citep{vonneumann_theory_1944}. The case of extensive-form games was initially problematic given that the corresponding formulation grows exponentially with respect to the number of states. The usual approach was to convert extensive-form games to normal-form and then solving the corresponding linear program.

In the nineties, Koller et al.\;provided an alternative (KMvS algorithm\citep{koller_fast_1994}) with much better scaling properties by representing strategies via \textit{realization weights}\citep{koller_complexity_1992}. The formulation results in linear programs are are proportional to the game tree size\citep{vonstengel_computing_2002}. An excellent review of these methods is provided by\citet{vonstengel_computing_2002}.

Dealing with uncertainty also brings its own challenges. Payoff uncertainty is common in many applications\citep{wang_network_2011}\citep{rass_defending_2017}\citep{lalropuia_modeling_2019}\citep{banks_adversarial_2015} and introduces additional problems.
For instance, Ravat et al.\;looked at stochastic Nash-Cournot games \citep{ravat_characterization_2011} instead of the more common deterministic variants. Recent research has explored other cases beyond payoffs such as players' strategies subject to stochastic constraints\citep{singh_second-order_2019}.

An initial classic approach to handle random payoffs is to solve a game that uses only the corresponding expected values\citep{pat-cornell_probabilistic_2002}. However, as argued by many authors, this is not always adequate \citep{cheng_random-payoff_2016}\citep{singh_characterization_2018}\citep{banks_adversarial_2015}\citep{singh_characterization_2018}. Harsanyi introduced the concept of disturbed games \citep{harsanyi_games_1973} where players' uncertainty of the exact payoff values is modeled by introducing random perturbations. One of the arguments against using expected values is that information on variability and uncertainty is not available to stakeholders that would like to assess the risk of certain strategies \citep{singh_characterization_2018}. A precise equilibrium calculation using the expected value of an uncertain belief model can be argued as misleading and maintaining information about uncertainty can inform stakeholders on the stability and robustness of the proposed solutions.

Classically, a stochastic programming formulation is proposed for the special case of normal-form games with independent and normally distributed random variables in the payoff\citep{bhurjee_optimal_2017}. For discrete distributions, the formulation results in a mixed-integer linear programming (MILP) problem. A similar and related problem where payoffs are modeled by closed intervals is expressed as two linear programs that are solved to find the corresponding bounds\citep{bhurjee_optimal_2017}.

In general, most work on stochastic games concentrates on linear programming (LP) based formulations and classic stochastic optimization methods. This is also the preferred approach for stochastic cooperative games with random payoffs\citep{suijs_cooperative_1999}\citep{ozen_general_2009}\citep{uhan_stochastic_2015}. Even without considering uncertainty and stochasticity, LP-based methods rapidly hit some limits as game complexity increases. Even with techniques to find isomorphisms in the extensive form\citep{gilpin_optimal_2005}, linear models for games such as Poker rapidly reach computational or memory limits and become intractable.

From a Bayesian game perspective, games with random payoff functions can be restructured and modeled through a probability distribution over a continuous player type space. However, in practice this is not as simple as it seems. Rabinovich proposed an approach based on fictitious-play\citep{brown_iterative_1951} to solve one-shot symmetric games with continuous single-dimensional types where utilities must be independent from adversary types\citep{rabinovich_computing_2013}.

Our work addresses Bayesian games with continuous types arising from epistemic uncertainty by exploiting some of the recent advances in Poker research and extrapolating them to other applications. In particular, we explore the problem of arbitrary continuous payoff distributions in two-player zero-sum extensive-form games.

\subsection{Poker and Artificial Intelligence}
Large-scale perfect information games have been a long standing challenge for Artificial Intelligence research. Recent advances have resulted in superhuman performance in chess, shogi, and Go\citep{silver_general_2018}.

Unlike perfect information games, the strategic implications of information asymmetry significantly increases complexity. For this reason, imperfect information games have received attention, and recent developments have concentrated on real-life variations of Poker, a challenging large-scale problem. Linear models have been conventionally used: most poker agents in the first Computer Poker Competition (2007) used LP formulations and packages\citep{waugh_unified_2014}.

Counterfactual Regret Minimization (CFR) has been a breakthrough in the effort to solve larger games\citep{zinkevich_regret_2008}\citep{sandholm_solving_2015}. Progress has been exciting in the last few years. Heads-up limit hold'em (HULHE), an imperfect information game with $3.19 \cdot 10^{14}$ decision points, has been weakly-solved\citep{bowling_heads-up_2015}. Moreover, algorithms for Heads-up no-limit hold'em (HUNL), a more challenging variant with more than $10^{160}$ decision points, have achieved superhuman performance\citep{moravcik_deepstack_2017}\citep{brown_superhuman_2018} beating professional players with statistical significance.

Since CFR was proposed, its success has resulted in derivative work improving and building upon this idea. In particular, variants such as MCCFR \citep{lanctot_monte_2013}, CFR+\citep{tammelin_solving_2014}, or LCFR/DCFR \citep{brown_solving_2018} have been progressively achieving better results. Another interesting line of research has focused on combining CFR with function approximation techniques such as neural networks \citep{brown_deep_2018}\citep{li_double_2018} to reduce even further the memory requirements of large game trees.

In this paper, for the sake of simplicity, we will limit the discussion to the most basic CFR formulation, typically referred to as \textit{Vanilla CFR}. Nevertheless, we expect our contribution to be easily extended to newer developments in the CFR research agenda.

% %%%%%%%%%%%%%%%%%%%%%%%%%%%%%%%%%%%%%%%%%%%%%%
% %%%%%%%%%%%%%%%%%%%%%%%%%%%%%%%%%%%%%%%%%%%%%%
% %%%%%%%%%%%%%%%%%%%%%%%%%%%%%%%%%%%%%%%%%%%%%%

\section{Background}

The following section introduces two-player games of imperfect information and discusses $\varepsilon$-Nash equilibrium. Moreover, we provide a short introduction on regret matching and counterfactual regret minimization. We also introduce definitions and notation adapted from \citet{shoham_multiagent_2008}, \citet{zinkevich_regret_2008}, \citet{lanctot_monte_2013}, and  \citet{gibson_regret_2014}.

\subsection{Imperfect information games (extensive-form)}
A two-player zero-sum extensive-form game of imperfect information $\Gamma$ is defined by a tuple $(\mathcal{N}, \mathcal{A}, \mathcal{H}, \mathcal{Z}, A, P, s, u, \mathcal{I})$ \citep{shoham_multiagent_2008}\citep{gibson_regret_2014} where:
\begin{align*}
    \mathcal{N} & = \{1, 2\}                                                                &  & \text{set of players}                     \\
    \mathcal{A} &                                                                           &  & \text{set of actions}                     \\
    \mathcal{H} &                                                                           &  & \text{set of non-terminal histories}      \\
    \mathcal{Z} &                                                                           &  & \text{set of terminal histories}          \\
    A           & : \mathcal{H} \rightarrow 2^{\mathcal{A}}                                 &  & \text{available actions function}         \\
    P           & : \mathcal{H} \rightarrow \mathcal{N}                                     &  & \text{player function that assigns turns} \\
    s           & : \mathcal{H} \times \mathcal{A} \rightarrow \mathcal{H} \cup \mathcal{Z} &  & \text{successor function, game tree}      \\
    u           & : \mathcal{N} \times \mathcal{Z} \rightarrow \mathbb{R}                   &  & \text{utility function}                   \\
    \mathcal{I} & : (\mathcal{I}_1, \mathcal{I}_2)                                          &  & \text{a tuple of information partitions}
\end{align*}
For zero-sum games, $u_1(z) = -u_2(z)$ for any given terminal state $z$. If a game involves chance, a third special player $\mathfrak{N}$ --- typically referred to as Nature --- should be considered and $\mathcal{N}= \{1, 2\} \cup \mathfrak{N}$. 

Histories $h \in \mathcal{H} \cup \mathcal{Z}$ are sequences of actions. All histories start with $\varnothing$ (empty history). The game finishes at terminal histories $z \in \mathcal{Z}$ that have no available actions, and payoffs are awarded at this point. The prefix relation between histories is denoted with $h \sqsubseteq h'$ and indicates that it is possible to obtain  $h'$ by adding actions to the sequence $h$. Lastly, the subset of histories where $i$ makes decisions is $\mathcal{H}_i = \{ h \in \mathcal{H} | P(h)=i \}$\citep{gibson_regret_2014}. 

There are states where player $i$ lacks the data to recognize the difference between two independent histories. This is represented by the information partition $\mathcal{I}_i$ that partitions $\mathcal{H}_i$ into information sets $I_{i,k}$. To denote that information set $I$ is a prefix of $h$, we will use $I \sqsubseteq h$. We also use $\mathcal{Z}_{I}$ for the subset of terminal histories $z \in \mathcal{Z}$ where $I \sqsubseteq z$.

The strategy that player $i \in \mathcal{N}$ follows is denoted by $\sigma_i(a|I_{i,k}): \mathcal{I}_i \times \mathcal{A} \rightarrow [0,1]$ that assigns a probability to each action $a$ given an information set $I_{i,k}$.  Given an action, the successor function $s$ is applied resulting in a new state. This repeats until the game reaches a terminal state $z$.  A strategy profile $\sigma$ is a tuple of the strategies of all players. We indicate player $i$ with a subscript $i$, and we use $-i$ to indicate all players except $i$. Also $\sigma_{I \mapsto a}$ denotes a strategy identical to $\sigma$ except from $I$ where $\sigma(a|I)=1$.

Given a history $h$, the active player at that point is $P(h)$, and the corresponding strategy can be expressed as $\sigma(h)=\sigma_{P(h)}(h)$. When all players follow $\sigma$, the probability of history $h$ is given by its \textit{reachability}:
\begin{align}
    \pi^\sigma(h) = \prod\limits_{ s(x,a) \sqsubset h} \sigma(a|x)
\end{align}
This can be extended to information sets:
\begin{align}
    \pi^\sigma(I) = \sum\limits_{h \in I} \pi^\sigma(h)
\end{align}
The expected payoff strategy $\sigma$ is:
\begin{align}
    u_i(\sigma) = \mathbb{E}_{\sigma} u_i(z) = \sum\limits_{z\in \mathcal{Z}} \pi^\sigma(z)\;u_i(z)
\end{align}

Players try to maximize their expected utility by adjusting their strategies. An $\varepsilon$-Nash generalizes Nash equilibrium. In this case, given other players' strategies, player $i$ can only improve its payoff by a small value $\varepsilon$. We will refer to Nash-equilibrium strategies as $\sigma^*$, and equally, we will use $\sigma^{*\varepsilon}$ for $\varepsilon$-Nash. For an $\varepsilon = 0$, the definition reduces to a classical Nash equilibrium. This can be formally expressed as:
\begin{align}
    u_i(\sigma) + \varepsilon \geq \max\limits_{\sigma'_i} u_i(\sigma'_i, \sigma_{-i}) \; \forall i \in \mathcal{N}
\end{align}

Along this work, \textit{perfect recall} is assumed, i.e., players remember complete histories making them equivalent to states. This assumption helps to maintain our presentation simple. Nevertheless, recent research has explored and analyzed convergence bound guarantees to a wider set of games that do not require this assumption \citep{gibson_regret_2014}. We plan to consider these ideas in future research.

\subsection{Regret matching}
\citet{hart_simple_2000} proposed regret matching (RM) as a way to minimize cumulative regret for normal-form games. As players iteratively try to learn an optimal strategy, they may try different strategies $\sigma^t$ over time $1 \leq t \leq T$. At any $t$, if player $i$ could replace all its previous strategies with some optimal $\sigma_i^{*t}$, the corresponding cumulative regret would be:
\begin{align}
    R_i^{t} = \max\limits_{\sigma^{*t}_i} \sum\limits^n_{k=1}\Big[ u_i(\sigma^{*t}_i, \sigma^k_{-i}) - u_i(\sigma^k)\Big]
\end{align}
similarly, this can also be defined at the action level:
\begin{align}
    R_i^{t}(a) = \sum\limits^n_{k=1}\Big[ u_i(a, \sigma^k_{-i}) - u_i(\sigma^k)\Big]
\end{align}
that is, regretting not choosing a fixed action $a$ instead of the corresponding $\sigma^k_i$ used in the past.

The method iteratively approximates correlated equilibria \citep{aumann_subjectivity_1974}, and in the particular case of constant-sum games, the resulting strategies coincide with an $\varepsilon$-Nash equilibrium \citep{lanctot_monte_2013}\citep{gibson_regret_2013}. Regret matching tracks the current cumulative regret $R_i^t(a)$ for each action $a$ and uses a simple update rule for strategies.
We will denote $x^+=\max\{x,0\}$ to truncate negative values, and similarly, we will use $R_i^{t, +}(a)$ to truncate negative regrets to zero.
At each iteration $t$, strategies are updated using:
\begin{align}
    \sigma_i^{t+1}(a) = \frac{R_i^{t, +}(a) }{\sum_{x} R_i^{t, +}(x)}\label{eq:update_rm}
\end{align}
When the denominator $R_i^{t, +}(a) = 0$, the classic approach \citep{zinkevich_regret_2008} is to use $\sigma_i^{t+1}(a) = |\mathcal{A}|^{-1}$ (uniform distribution over actions), however, recent work \citep{brown_deep_2018} opts for selecting the action with the highest $R_i^{t, +}$.

\subsection{Counterfactual regret minimization}

Counterfactual regret minimization (CFR) \citep{zinkevich_regret_2008}\citep{gibson_regret_2014} is an anytime learning algorithm that through iterative self-play approximates $\varepsilon$-Nash equilibrium and extends to two-player zero-sum extensive-form games.

A significant advantage is that regret minimization is done at each information set instead of at each state. The algorithm can traverse the game using the successor function $s$ and does not require storing the game tree, instead, information must be kept only for information sets so memory requirements are the order of $O(\sum_i |\mathcal{I}_i|)$\citep{lanctot_monte_2013}.

Counterfactual value is the key concept that extends regret to extensive-form games. This is defined as the expected payoff of information set $I$ taking into account reachabilities:
\begin{align}
    v_i(I, \sigma) = \sum\limits_{z \in \mathcal{Z}_I} u_i(z)
    ~\pi_{-i}^\sigma(h_{I, z})
    ~\pi^\sigma(z \mid h_{I, z})
\end{align}
where we use $h_{I, z}$ to indicate a history that both $h \in I$ and $h \sqsubseteq z$. Using the following definition, the counterfactual regret for the extended form is given by:
\begin{align}
    r^t_i(I, a) = v_i(I,\sigma^t_{I \mapsto a}) - v_i(I,\sigma^t)
\end{align}
The cumulative counterfactual regret is then:
\begin{align}
    R_i^t(I, a) = \sum\limits_{k=1}^{t} r_i^t(I,a)
\end{align}
and strategies are updated in a similar fashion as RM (eq. \ref{eq:update_rm}) but extended to information sets:
\begin{align}
    \sigma_i^{t+1}(I, a) = \frac{R_i^{t, +}(I, a) }{\sum_{x} R_i^{t, +}(I, x)}
\end{align}
At any point the average strategy $\bar\sigma$ can be calculated as:
\begin{align}
    \bar\sigma^T_i(I) = \frac{\sum^t_{k=1}\pi_i^{\sigma^k}(I) \; \sigma^t(I, a)}
    {\sum^t_{k=1} \pi_i^{\sigma^k}(I)}
\end{align}
In practice, there is no need to keep previous strategies \citep{gibson_regret_2014}, instead, a running sum $S$ is kept:
\begin{align}
    S^{t+1}_i(I, a) = S^{t}_i(I, a) + \pi_i^{\sigma^k}(I) \; \sigma(I,a)
\end{align}
that later can be used to calculate the average strategy by normalizing:
\begin{align}
    \bar\sigma^T_i(I) =
    \frac{S^T_i(I,a)}
    {\sum_{x \in A(I)} S^T_i(I, x) }
\end{align}

Vanilla CFR does not make any particular consideration for chance nodes (Algorithm \ref{alg:vanilla}), and in each iteration all information sets and applicable actions are explored.
\begin{algorithm}
    \caption{Vanilla CFR (adapted from \citet{gibson_regret_2014})}\label{alg:vanilla}
    \SetAlgoLined{}
    \SetNoFillComment{}
    \SetKwBlock{Init}{Initialization:}{}
    \Init{
    \tcp*[h]{set accumulators to zero} \\
    $R(I,a) \leftarrow 0$ \\
    $S(I,a) \leftarrow 0$ \\
    \tcp*[h]{initialize to uniform distribution} \\
    $\sigma(I,a) \leftarrow |\mathcal{A}|^{-1}$ \\
    }
    \
    \For{t in $1 \ldots T$}{
        \For{i in $\mathcal{N}$}{
            \For{I in $\mathcal{I}_i$}{
                $\sigma(I, \cdot) \leftarrow$ RegretMatching$(\;R(I, \cdot)\;)$     \\
                \For{a in $A(I)$}{
                    $R(I, a) \leftarrow R(I,a)+ v_i(I, \sigma_{I \mapsto a}) - v_i(I, \sigma)$                \\
                    $S(I, a) \leftarrow S(I,a) + \pi_i^\sigma(I)\;\sigma_i(I,a)$
                }
            }
        }
    }
    \
    Normalize $S(I)$
\end{algorithm}
%%%%%%%%%%%%%%%%%%%%%%%%%%%%%%%%%%%%%%%%%%%%%%
%%%%%%%%%%%%%%%%%%%%%%%%%%%%%%%%%%%%%%%%%%%%%%
%%%%%%%%%%%%%%%%%%%%%%%%%%%%%%%%%%%%%%%%%%%%%%
\section{Harsanyi-Counterfactual Regret Minimization (H-CFR)}
When dealing with strategic conflict, we are interested in allowing for complex and rich payoff models that can not only be solved but also provide additional risk related information to stake holders. Instead of formulating the problem as a linear program, in order to deal with arbitrary payoff models, we transform the game with a simple modification and proceed to solve it using counterfactual regret minimization. 

The Harsanyi transformation \citep{harsanyi_games_1967} converts an incomplete game into an imperfect game. Given a game $\Gamma$ with one or many continuous random functions $\mathcal{P}$ modeling payoffs, we can generate a modified game $\Gamma^{\#}$ by prepending a single node $\mathbb{H}$ as root node where all payoff functions for the complete tree are sampled. This new root leads to the original root, now representing a very large information set $I_{\Gamma}$ where each $\mathbb{H}$ outcome results in a complete game tree $\Gamma$.
\begin{figure}
    \centering
    \def\svgwidth{9cm}
%% Accompanies image file 'tree.pdf' (pdf, eps, ps)
\begingroup%
  \makeatletter%
  \providecommand\color[2][]{%
    \errmessage{(Inkscape) Color is used for the text in Inkscape, but the package 'color.sty' is not loaded}%
    \renewcommand\color[2][]{}%
  }%
  \providecommand\transparent[1]{%
    \errmessage{(Inkscape) Transparency is used (non-zero) for the text in Inkscape, but the package 'transparent.sty' is not loaded}%
    \renewcommand\transparent[1]{}%
  }%
  \providecommand\rotatebox[2]{#2}%
  \newcommand*\fsize{\dimexpr\f@size pt\relax}%
  \newcommand*\lineheight[1]{\fontsize{\fsize}{#1\fsize}\selectfont}%
  \ifx\svgwidth\undefined%
    \setlength{\unitlength}{656.1293276bp}%
    \ifx\svgscale\undefined%
      \relax%
    \else%
      \setlength{\unitlength}{\unitlength * \real{\svgscale}}%
    \fi%
  \else%
    \setlength{\unitlength}{\svgwidth}%
  \fi%
  \global\let\svgwidth\undefined%
  \global\let\svgscale\undefined%
  \makeatother%
  \begin{picture}(1,0.30122598)%
    \lineheight{1}%
    \setlength\tabcolsep{0pt}%
    \put(0,0){\includegraphics[width=\unitlength,page=1]{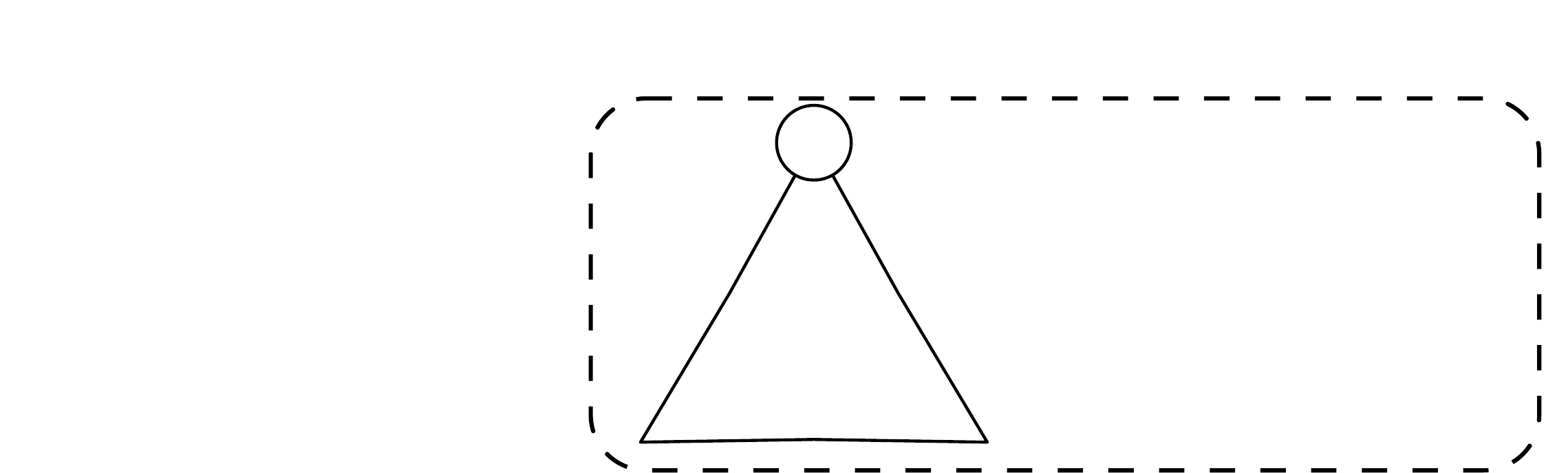}}%
    \put(0.51242698,0.08343413){\color[rgb]{0,0,0}\makebox(0,0)[t]{\smash{\begin{tabular}[t]{c}$\Gamma_1$\end{tabular}}}}%
    \put(0,0){\includegraphics[width=\unitlength,page=2]{tree.pdf}}%
    \put(0.83981431,0.08343413){\color[rgb]{0,0,0}\makebox(0,0)[t]{\smash{\begin{tabular}[t]{c}$\Gamma_\infty$\end{tabular}}}}%
    \put(0.67732908,0.08343413){\color[rgb]{0,0,0}\makebox(0,0)[t]{\smash{\begin{tabular}[t]{c}$\ldots$\end{tabular}}}}%
    \put(0.66886207,0.19809761){\color[rgb]{0,0,0}\makebox(0,0)[t]{\smash{\begin{tabular}[t]{c}$I_{\mathbb{H}}$\end{tabular}}}}%
    \put(0,0){\includegraphics[width=\unitlength,page=3]{tree.pdf}}%
    \put(0.10475833,0.08343413){\color[rgb]{0,0,0}\makebox(0,0)[t]{\smash{\begin{tabular}[t]{c}$\Gamma$\end{tabular}}}}%
    \put(0,0){\includegraphics[width=\unitlength,page=4]{tree.pdf}}%
    \put(0.67230328,0.25925207){\color[rgb]{0,0,0}\makebox(0,0)[t]{\smash{\begin{tabular}[t]{c}$\mathbb{H}$\end{tabular}}}}%
  \end{picture}%
\endgroup%
\label{fig:harsanyi}
    \caption{Transformation from $\Gamma$ to $\Gamma^{\#}$}
\end{figure}
We later apply CFR \citep{zinkevich_regret_2008} to the modified game $\Gamma^{\#}$. Instead of solving several linear programs as \citet{cheng_random-payoff_2016} do, we iteratively approximate a solution. Our approach allows approximating equilibria for any arbitrary random function without the restrictions introduced by other classic stochastic optimization based proposals and can be applied to extensive-form game when the number of states is large.
\begin{algorithm}
    \caption{H-CFR:Harsanyi-Counterfactual Regret Minimization}\label{alg:hcfr}
    \SetAlgoLined{}
    \SetNoFillComment{}
    \SetKwBlock{Init}{Initialization:}{}
    \SetKwBlock{Transform}{H-Transform:}{}
    \SetKwBlock{Solve}{Solve:}{}

    \KwData{$\Gamma$, $T$}
    \KwResult{$\sigma_{*\varepsilon}$}

    \Transform{
        Link $\mathbb{H}$ to all nodes $z \in \mathcal{Z}$ with random variables modeling uncertainty.\\
        Set old root information state $I_\mathbb{H}$ to $\Gamma_{root}$ \\
        $\Gamma^\# \leftarrow$ Prepend $\mathbb{H}$ to $\Gamma_{root}$ \\
    }
    \
    \Solve{
        $\sigma^{*\varepsilon} \leftarrow$ CFR$( \Gamma^\#, T )$
    }
\end{algorithm}
The number of iterations $T$ for CFR to find an $\varepsilon$-Nash equilibrium is bounded by $O(\varepsilon^{-2})$ \citep{lanctot_monte_2013}, however, the learning process depends on information retrieved by playing the game. For this reason, it is arguable that different $\mathcal{P}$ distributions may affect the required number of iterations due to sampling issues.
Even further, while in a different context, related research \citep{lanctot_monte_2013} has analyzed the use of importance sampling and some of these results could be transferred to
this line of research. We intend to explore them in future research.
%%%%%%%%%%%%%%%%%%%%%%%%%%%%%%%%%%%%%%%%%%%%%%
%%%%%%%%%%%%%%%%%%%%%%%%%%%%%%%%%%%%%%%%%%%%%%
%%%%%%%%%%%%%%%%%%%%%%%%%%%%%%%%%%%%%%%%%%%%%%
%%%%%%%%%%%%%%%%%%%%%%%%%%%%%%%%%%%%%%%%%%%%%%
\subsection{Numerical experiments}
\subsubsection{Simple routing problem (extensive-form)}
\begin{figure}
    \centering
    \def\svgwidth{9cm}
%% Accompanies image file 'routing.pdf' (pdf, eps, ps)
\begingroup%
  \makeatletter%
  \providecommand\color[2][]{%
    \errmessage{(Inkscape) Color is used for the text in Inkscape, but the package 'color.sty' is not loaded}%
    \renewcommand\color[2][]{}%
  }%
  \providecommand\transparent[1]{%
    \errmessage{(Inkscape) Transparency is used (non-zero) for the text in Inkscape, but the package 'transparent.sty' is not loaded}%
    \renewcommand\transparent[1]{}%
  }%
  \providecommand\rotatebox[2]{#2}%
  \newcommand*\fsize{\dimexpr\f@size pt\relax}%
  \newcommand*\lineheight[1]{\fontsize{\fsize}{#1\fsize}\selectfont}%
  \ifx\svgwidth\undefined%
    \setlength{\unitlength}{346.35468106bp}%
    \ifx\svgscale\undefined%
      \relax%
    \else%
      \setlength{\unitlength}{\unitlength * \real{\svgscale}}%
    \fi%
  \else%
    \setlength{\unitlength}{\svgwidth}%
  \fi%
  \global\let\svgwidth\undefined%
  \global\let\svgscale\undefined%
  \makeatother%
  \begin{picture}(1,0.32736912)%
    \lineheight{1}%
    \setlength\tabcolsep{0pt}%
    \put(0,0){\includegraphics[width=\unitlength,page=1]{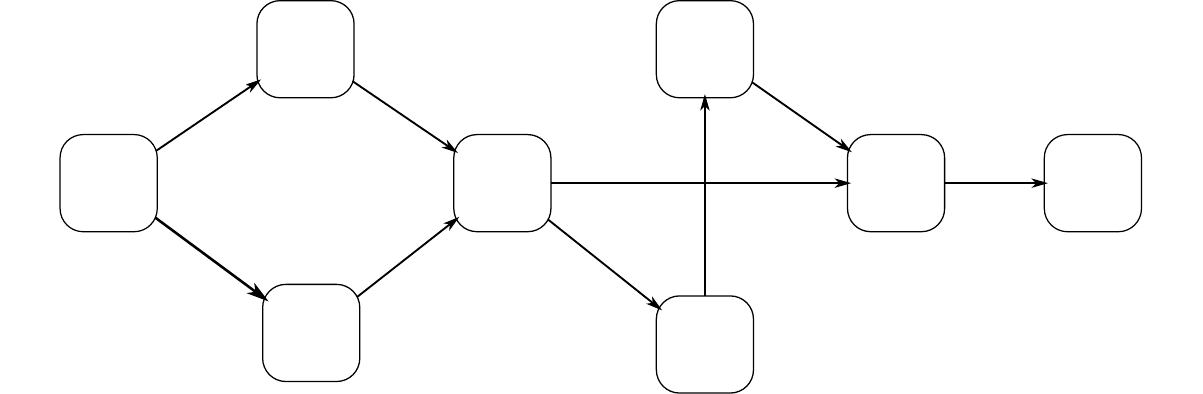}}%
    \put(0.09031566,0.16710863){\color[rgb]{0,0,0}\makebox(0,0)[t]{\smash{\begin{tabular}[t]{c}$\mathbb{S}$\end{tabular}}}}%
    \put(0.25400022,0.27847712){\color[rgb]{0,0,0}\makebox(0,0)[t]{\smash{\begin{tabular}[t]{c}$v_1$\end{tabular}}}}%
    \put(0.90873839,0.16710863){\color[rgb]{0,0,0}\makebox(0,0)[t]{\smash{\begin{tabular}[t]{c}$\mathbb{T}$\end{tabular}}}}%
    \put(0.25875092,0.04254966){\color[rgb]{0,0,0}\makebox(0,0)[t]{\smash{\begin{tabular}[t]{c}$v_2$\end{tabular}}}}%
    \put(0.41768478,0.16710863){\color[rgb]{0,0,0}\makebox(0,0)[t]{\smash{\begin{tabular}[t]{c}$v_3$\end{tabular}}}}%
    \put(0.58611995,0.27847712){\color[rgb]{0,0,0}\makebox(0,0)[t]{\smash{\begin{tabular}[t]{c}$v_4$\end{tabular}}}}%
    \put(0.58611995,0.03295028){\color[rgb]{0,0,0}\makebox(0,0)[t]{\smash{\begin{tabular}[t]{c}$v_5$\end{tabular}}}}%
    \put(0.74505396,0.16710863){\color[rgb]{0,0,0}\makebox(0,0)[t]{\smash{\begin{tabular}[t]{c}$v_6$\end{tabular}}}}%
  \end{picture}%
\endgroup%

    \caption{Routing problem, based on \citet{wang_network_2011}}\label{fig:problem_wang}
\end{figure}

As an example, we address the routing problem proposed by \citet{wang_network_2011}. A model for a two-player simultaneous game: attacker ($\mathbb{A}$) vs defender ($\mathbb{D}$). Given a graph such as that in Figure \ref{fig:problem_wang}, the defender chooses a route from $\mathbb{S}$ to $\mathbb{T}$. The attacker $\mathbb{A}$ decides in advance which nodes $v_i$ to booby-trap with an improvised explosive device (IED). The defender's convoy traverses a route accumulating damage as it passes nodes that had been selected by the attacker. The game can be defined for any network and number of IEDs. The analysis is done from the defender's perspective that makes a subjective assessment of the possible damage at each node, summarized in the payoff $\mathcal{P}$.

\paragraph{Original approach}
\citet{wang_network_2011} solve the given example using a variation of fictitious play that they denominate adversarial risk analysis mirroring fixed point (ARA-MFP).

An initial restriction is that their approach requires a normal-form formulation so all possible valid routes need to be calculated in advance. As a result, Defender actions $r_i$ refer to each of these possible routes. In the example, the number of routes is actually small but can dramatically increase for other networks:
\begin{align}
    \begin{split}
        r_1 & : \mathbb{S} \rightarrow v_1 \rightarrow v_3 \rightarrow v_5 \rightarrow v_4 \rightarrow v_6 \rightarrow \mathbb{T} \\
        r_2 & : \mathbb{S} \rightarrow v_1 \rightarrow v_3 \rightarrow v_6 \rightarrow \mathbb{T}                                 \\
        r_3 & : \mathbb{S} \rightarrow v_2 \rightarrow v_3 \rightarrow v_5 \rightarrow v_4 \rightarrow v_6 \rightarrow \mathbb{T} \\
        r_4 & : \mathbb{S} \rightarrow v_2 \rightarrow v_3 \rightarrow v_6 \rightarrow \mathbb{T}
    \end{split}
\end{align}
The attacker $\mathbb{A}$ can use at most one IED, and the action space is indicated by the set of nodes or $\varnothing$ for no attack.
\citet{wang_network_2011} model the problem as a zero-sum game, and matrices define the outcome of each attacker/defender interaction:
\begin{align}
    \mathbf{U}_\mathbb{A} = -\mathbf{U}_\mathbb{D} =
    \bordermatrix{
                & r_1 & r_2 & r_3 & r_4 \cr
    \varnothing & 0   & 0   & 0   & 0 \cr
    v_1         & U_1 & U_1 & 0   & 0 \cr
    v_2         & 0   & 0   & U_2 & U_2 \cr
    v_3         & U_3 & U_3 & U_3 & U_3 \cr
    v_4         & U_4 & 0   & U_4 & 0 \cr
    v_5         & U_5 & 0   & U_5 & 0 \cr
    v_6         & U_6 & U_6 & U_6 & U_6 \cr
    } \qquad
\end{align}
\paragraph{Our proposal}
In H-CFR, it is possible to work directly with the extensive form of the game so there is no need to convert to a normal form. For each run, we refer to the set of visited nodes as $\mathcal{V}$, and $\bm{1}_\mathcal{V}$ is the corresponding indicator function for this set. The payoff at the terminal node $\mathbb{T} \in \mathcal{Z}$ can be then defined with a simple expression:
\begin{align}
    u_\mathbb{A}(\mathbb{T})= \sum_{k} U_k \cdot \bm{1}_\mathcal{V}(v_k)
\end{align}
As previously described, we transform the game tree $\Gamma$ into $\Gamma^\#$ by prepending a node $\mathbb{H}$ where payoff sampling occurs, and we approximate the $\varepsilon$-Nash equilibrium using CFR. This can be easily extended to other more advanced CFR variations.

\paragraph{Binomial distribution}
The original problem models interactions as independent and equal to $U_k = Binomial(10, 0.5)$. Based on these parameters, Table \ref{tbl:solutions_wang2} shows results obtained for both methods.
\begin{table}
    \caption{Solutions --- Attacker's strategy ($\sigma_\mathbb{A}(a)$)}\label{tbl:solutions_wang2}
    \centering
    \begin{tabular}{ccc}
        \toprule
        \cmidrule(r){1-2}
                      & ARA-MFP  & H-CFR (binomial) \\
        \midrule
        $iterations$  & $3598$   & $500$            \\
        $\varnothing$ & $0.0000$ & $0.0003$         \\
        $v_1$         & $0.1241$ & $0.0003$         \\
        $v_2$         & $0.1241$ & $0.0011$         \\
        $v_3$         & $0.4275$ & $0.5376$         \\
        $v_4$         & $0.0001$ & $0.0003$         \\
        $v_5$         & $0.0000$ & $0.0005$         \\
        $v_6$         & $0.3243$ & $0.4602$         \\
        \bottomrule
    \end{tabular}
\end{table}

Given the specific routing network selected in the example, an infinite number of equilibria exist for both players. While it could be expectable that our equilibrium does not match that of Wang, when analyzed in detail, we recognize that the published result is far from a Nash equilibrium.

Given the network structure, the attacker will always damage the defender by placing an IED in node $v_3$ or $v_6$. Surprisingly, the solution published by \citet{wang_network_2011} recommends $v_1$ or $v_2$ as part of the strategy, resulting in a damage probability significantly lower than 1.

Our method (H-CFR) effectively converges to an adequate strategy. Moreover, we emphasize that the algorithm  works directly with the extensive form of the game and requires minimal changes to the problem definition. Also, much more interesting payoff models can be used. In particular, we intend to explore games with payoffs represented as Bayesian networks in future research.

\begin{figure}
    \centering
    \input{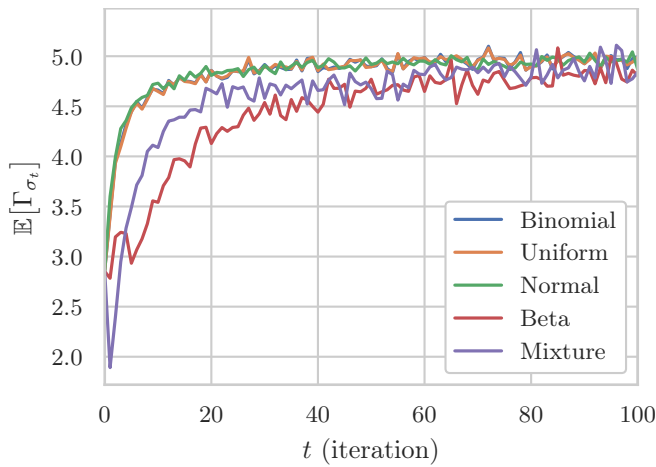}
    \captionsetup{justification=centering}
    \caption{Expected value convergence for different payoff models}\label{fig:exp_value}
\end{figure}

\begin{figure}
    \centering
    \input{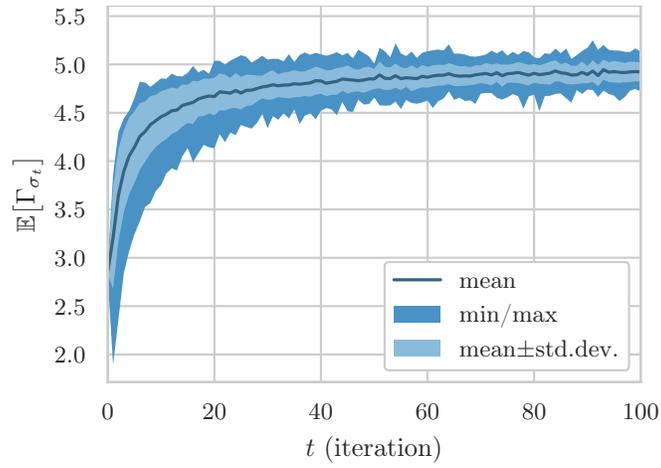}
    \captionsetup{justification=centering}
    \caption{Expected value convergence for the mixture model}\label{fig:exp_value_mixture}
\end{figure}

\begin{figure}
    \centering
    \input{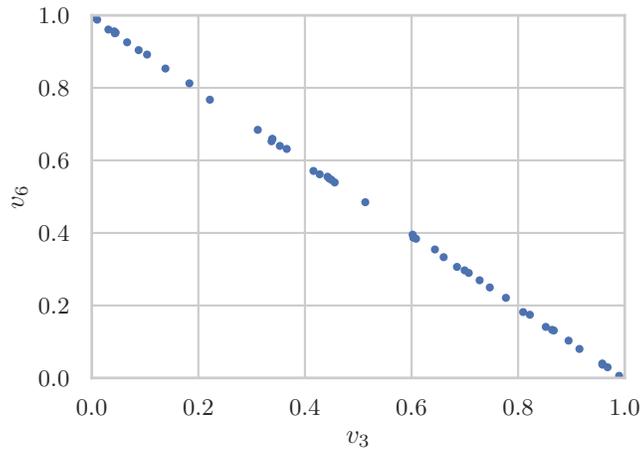}
    \captionsetup{justification=centering}
    \caption{Strategy values for $v_3$ and $v_6$ for the mixture model}\label{fig:equilibria_mix}
\end{figure}

\begin{figure}
    \centering
    \input{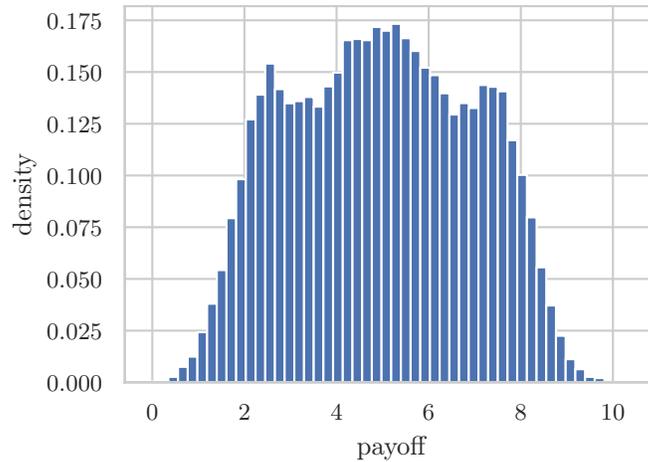}
    \captionsetup{justification=centering}
    \caption{Payoff distribution for equilibrium strategy (mixture model)}\label{fig:samples_mix}
\end{figure}

\paragraph{Other distributions}
We extend the payoff model to other probability distributions. Table \ref{tbl:models} provides more information on the specific parameters. All models share the same expected value to simplify comparisons.
\begin{table}
    \caption{Payoff models}\label{tbl:models}
    \centering
    \begin{tabular}{ll}
        \toprule
        \cmidrule(r){1-2}
                 & Model definition                            \\
        \midrule
        Binomial & $U_i=$Binomial$(0.5, 10)$                   \\
        Normal   & $U_i=$Normal$(5, 1)$                        \\
        Uniform  & $U_i=$Uniform$(0.5, 10)$                    \\
        Beta     & $U_i=$10 $\cdot $Beta$(0.5, 0.5)$           \\
        Mixture  & $U_i= $Normal$(2.5, 1)$ or Normal$(7.5, 1)$ \\
        \bottomrule
    \end{tabular}
\end{table}
We apply H-CFR for 500 iterations to each case. The corresponding approximations are shown in Table \ref{tbl:solutions_manydist}, and Figure \ref{fig:exp_value} shows the convergence of $\mathbb{E}[\Gamma_{\sigma_t}]$ with respect to iterations $t$.
\begin{table}
    \centering
    \captionsetup{justification=centering}
    \caption{Solutions --- Attacker's strategy for different belief models ($\sigma_\mathbb{A}$)\\ (500 iterations)}\label{tbl:solutions_manydist}
    \begin{tabular}{cccccc}
        \toprule
                      & Binomial & Uniform  & Normal   & Beta     & Mixture  \\
        \midrule
        $\varnothing$ & $0.0003$ & $0.0003$ & $0.0003$ & $0.0003$ & $0.0003$ \\
        $v_1$         & $0.0003$ & $0.0003$ & $0.0003$ & $0.0007$ & $0.0037$ \\
        $v_2$         & $0.0011$ & $0.0004$ & $0.0011$ & $0.0014$ & $0.0036$ \\
        $v_3$         & $0.5376$ & $0.6646$ & $0.3440$ & $0.9903$ & $0.9847$ \\
        $v_4$         & $0.0003$ & $0.0003$ & $0.0003$ & $0.0003$ & $0.0003$ \\
        $v_5$         & $0.0005$ & $0.0003$ & $0.0003$ & $0.0005$ & $0.0003$ \\
        $v_6$         & $0.4602$ & $0.3339$ & $0.6539$ & $0.0065$ & $0.0071$ \\
        \bottomrule
    \end{tabular}
\end{table}
While simpler models such as Binomial, Uniform, or Normal rapidly approach 5.0, Beta or Mixture models result in higher variance and require a higher number of iterations.
A general bound for CFR has been proven \citep{zinkevich_regret_2008}\citep{lanctot_monte_2013} so convergence to an $\varepsilon$-Nash in the limit is ensured.
We experimentally explored the case of a mixture model by running the same procedure multiple times (n=50).
Figure \ref{fig:exp_value_mixture} shows sampled extreme, average, and 1 standard error values for each iteration among all runs.
The plot shows how variance reduces over time as the correct strategy is being learned. However, as the value approaches convergence, there is an intrinsic sampling error due to the structure of the game. Figure \ref{fig:equilibria_mix} shows how we sample different comparable equilibria over different runs. In this case, resulting strategies indicate a linear combination of $v_3$ and $v_6$. These strategies can also be evaluated from a risk perspective by sampling (Figure \ref{fig:samples_mix}) to evaluate the probability of achieving certain minimum payoff.
%%%%%%%%%%%%%%%%%%%%%%%%%%%%%%%%%%%%%%%%%%%%%%
%%%%%%%%%%%%%%%%%%%%%%%%%%%%%%%%%%%%%%%%%%%%%%
%%%%%%%%%%%%%%%%%%%%%%%%%%%%%%%%%%%%%%%%%%%%%%
%%%%%%%%%%%%%%%%%%%%%%%%%%%%%%%%%%%%%%%%%%%%%%
\section{Conclusions}
We have proposed H-CFR, an alternative to approach $\varepsilon$-Nash equilibria in zero-sum extensive-form games where payoff uncertainty can be modeled by arbitrary distributions.
In addition, we have provided numerical examples by applying the method to previously published models.

The method not only profits from better scalability due to CFR-based advantages but also requires limited changes in the formulation of the problem. It can be applied both to normal and extensive forms, it is an anytime algorithm, and does not constrain the kind of distributions used to model payoff uncertainty.

The current work lays out a series of questions that we plan to consider in future research.
In particular, we would like to explore the effect of different payoff models over convergence bounds and a better use of sampling techniques such as importance sampling. Most games will typically have multiple equilibria. Each H-CFR run converges to a single strategy that depends stochastically on the sampled payoffs. Figure \ref{fig:equilibria_mix} showed that is possible to sample different equilibria. We plan to investigate improved approaches to sample the whole Nash equilibria space. Last but not least, we see the opportunity of applying this method to more complex Bayesian Network based models.

\bibliography{mylibrary}
\bibliographystyle{dcu}
\end{document}